\documentclass[column,showpacs,showkeys,preprint,numbers,amsmath,amssymb]{revtex4}
\usepackage{graphicx}
\usepackage{subfigure}
\begin{document}

\title{\bf Interface control of ferroelectricity in LaNiO$_{3}$-BaTiO$_{3}$ superlattices}

\author{Yin-Zhong Wu$^{1,3,}$\footnote{Electronic address: yzwu@cslg.edu.cn}, Hai-Shuang Lu$^2$, Tian-Yi Cai$^2$ \& Sheng Ju$^{2,}$\footnote{Electronic address: jusheng@suda.edu.cn}}
\affiliation{ $^{1}$Jiangsu Laboratory of Advanced Functional Materials and Physics Department, Changshu Institute of Technology, Changshu
215500, P. R. China, $^{2}$Department of Physics and Jiangsu Key Laboratory of Thin Films, Soochow University, Suzhou 215006, P. R.
China,$^{3}$School of Mathematics and Physics, Suzhou University of Science and Technology, Suzhou 215009, P. R. China}

\vspace{20cm}

\begin{abstract}
LaNiO$_{3}$-BaTiO$_{3}$ superlattices with different types of interfaces are studied from first-principles density-functional theory. It is
revealed that the ferroelectricity in the superlattice with (NiO$_2$)$^-$/(BaO)$^0$ interfaces is enhanced from that of the superlattice with
(LaO)$^+$/(TiO$_2$)$^0$ interfaces. The origin lies at the polar discontinuity at the interface, which makes the holes localized within the
(NiO$_2$)$^-$/(BaO)$^0$ interface, but drives a penetration of electrons into BaTiO$_3$ component near (LaO)$^+$/(TiO$_2$)$^0$ interface. Our
calculations demonstrate an effective avenue to the robust ferroelectricity in BaTiO$_3$ ultrathin films.
\end{abstract}


\maketitle
\section{Introduction}
Ferroelectric ultrathin films and heterostructure have attracted great interest in recent years$^{1-14}$. The ferroelectricity, in particular,
the critical thickness, was found depend strongly on the proper choice of the interface. By controlling the chemical bonding at the interface,
both the enhanced and decreased ferroelectricity have been found in BaTiO$_3$ and PbTiO$_3$ ultrathin films. In fact, polar discontinuity at the
heterointerface widely exists at oxide based heterostructures$^{15-20}$. Generally, it will lead to carrier doping in formerly insulating
component and consequently emergent electronic phase, e.g., conducting two-dimensional electron gas in LaAlO$_3$-SrTiO$_3$ heterointerface. For
ferroelectric materials, as studied in the BaTiO$_3$ bulk system$^{21}$, such a kind of carrier doping will strongly decrease the
ferroelectricity, equally for the electron doping and the hole doping. In the heterostructures or ferroelectric ultrathin films, however, the
situation is yet not clear. In all the previous studies of the ferroelectric ultrathin films$^{1-12}$, metallic electrode either of SrRuO$_3$
(with alternating (SrO)$^0$ and (RuO$_2$)$^0$ layers) or Pt shows no polar features.

Generally, as demonstrated in Fig.~1, the charge and potential diagrams in the metal-ferroelectrics superlattice can be simply described by a
screening model. The charge distribution of symmetric metal-ferroelectrics interfaces with normal metal, such as Pt and Au, is illustrated in
Fig.~1(a) with corresponding potential energy in Fig.~1(d). However, if a positively charged layer (e.g. the (LaO)$^+$ layer) is allowed at the
interface, then, screening electrons will accumulate at adjacent ferroelectric layer (see Fig.~1(b), here spontaneous polarization is set zero
for clarity), and an inhomogeneous electron doping will take place in the ferroelectric component. Taking the polarization into account, as
illustrated in Fig.~1(e), the potential profile is the superposition of them. On the other hand, for a negatively charged layer, similarly
asymmetric potential but with reversal sign is shown in Fig.~1(f). Such a picture suggests that the ferroelectric performance in the thin film
geometry should be the same for electron doping case and hole doping one, the situation of which is similar to the bulk system$^{21}$.

In order to check the validity of the above picture, we use LaNiO$_3$ as electrode with alternating charged layers of (LaO)$^+$ and
(NiO$_2$)$^-$. Perovskite oxide LaNiO$_3$ is structurally compatible with ferroelectric BaTiO$_3$, and ferroelectric thin films with LaNiO$_3$
electrodes were found to possess better fatigue properties compared with other metallic oxides$^{22,23}$. As shown in Fig.~2,
LaNiO$_3$-BaTiO$_3$ superlattices are proposed. Here, two types of polar discontinuities are clearly distinguished, namely
(LaO)$^+$/(TiO$_2$)$^0$ with positive charge and (NiO$_2$)$^-$/(BaO)$^0$ negative one, which correspond to the situation described in Fig.~1(b)
and (c), respectively. In the following, we will provide a systematic first-principles study of atomic and electronic structures in these
LaNiO$_3$-BaTiO$_3$ superlattices. Our calculations provide solid evidence of contrast between electron doping and hole doping in nanoscale
ferroelectrics.

\section{Systems and methods}
 Our $ab$ $initio$ calculations are performed using the full-potential projector-augmented wave (PAW) method$^{24}$,
as implemented in the Vienna $ab$ $initio$ simulation package (VASP)$^{25}$. They are based on local density-approximation, which can describe
both ferroelectric BaTiO$_3$ and electrode LaNiO$_3$ well$^{26}$. A plane-wave cutoff of 500 eV is used throughout, and atomic relaxations are
converged using $10\times 10\times 1$ Monkhorst-Pack k-point sampling of Brillouin zone until the maximum Hellman-Feynman force acting on each
atom is less than 0.01 eV/\AA. PAW potentials are used to describe the electron-ion interaction with 9 valence electrons for La
($5p^{6}5d^{1}6s^{2}$), 16 for Ni ($3p^63d^84s^2$), 10 for Ba ($5s^25p^{6}6s^{2}$), Ti ($3p^63d^24s^2$), and 6 for O ($2s^22p^4$). Superlattice
is (001) epitaxially grown on SrTiO$_3$ substrate with the in-plane lattice constant of 3.866 \AA. The superlattice structures we use are
[LaO-(NiO$_2$-LaO)$_{10}$/(TiO$_2$-BaO)$_m$-TiO$_2$] and [NiO$_2$-(LaO-NiO$_2$)$_{10}$/(BaO-TiO$_2$)$_m$-BaO], as shown in Fig.~2 for $m=3$. The
(LaO)$^+$/(TiO$_2$)$^0$ interface here is similar to (LaO)$^+$/(TiO$_2$)$^0$ in LaAlO$_3$-SrTiO$_3$ superlattice$^{18}$ and
(NbO$_2$)$^+$/(SrO)$^0$ interface in KNbO$_3$-SrTiO$_3$ superlattice$^{19}$. And the second type interface (NiO$_2$)$^-$/(BaO)$^0$ has not been
reported in the literature.

\section{Results and discussions}
 First, we study the inherent ferroelectric displacement and its evolution with the thickness of ferroelectric
layer. In Fig.~3(a)-(f) and 3(g)-(l), the relative cation-anion
displacements within both LaNiO$_3$ and BaTiO$_3$ are plotted. In
BaTiO$_3$, the spontaneous polarization is proportional to the
relative cation-anion displacements$^{3}$. Here, the existence of
ferroelectricity in ferroelectric barrier is determined by the
non-zero total polarization of the whole BaTiO$_3$ barrier$^{11}$.
Clearly, for (LaO)$^+$/(TiO$_2$)$^0$ interfaces, one can see that
the BaTiO$_3$ film is paraelectric (PE) when the thickness of
BaTiO$_3$ is smaller than six unit cells. On the other hand, for
the case of (NiO$_2$)$^-$/(BaO)$^0$ interfaces, the
ferroelectricity of BaTiO$_3$ film will be preserved down to the
thickness of four unit cells. The above results differ greatly
from the SrRuO$_3$-BaTiO$_3$-SrRuO$_3$ capacitor, where the
critical thickness of BaTiO$_3$ film with SrO/TiO$_2$ interfaces
is thinner than that of BaTiO$_3$ film with RuO$_2$/BaO
interfaces$^{9}$. The polarity-discontinuity at
metal-ferroelectric interfaces is therefore dominating. In
addition, the relative cation-anion displacements for $m=8$
approach the bulk value of BaTiO$_3$.

To gain a deeper insight into the above differences, electronic
structure of local density of state is plotted in Fig. 4 for the
superlattice (LaNiO$_3$)$_{10.5}$-(BaTiO$_3$)$_{3.5}$, where the
BaTiO$_3$ film stays on paraelectric(PE) state as illustrated in
Fig. 3, Thus, the effect of polar discontinuity on the DOS of
BaTiO$_3$ is investigated without consideration of the influence
of spontaneous polarization at the first step. The layer-projected
DOS are centrosymmetric for both two types of interfaces. For
(LaO)$^+$/(TiO$_2$)$^0$ interfaces in Fig. 4(a), the Fermi energy
is pushed into the bottom of conduction band, which is similar to
the positively charged (LaO)$^+$ doped in
SrRuO$_3$-BaTiO$_3$-SrRuO$_3$ junctions$^6$. Due to the
discontinuity of polarity, electrons transfer from the metal to
BaTiO$_3$ film to neutralize the polarity, and charge leakage will
suppress polarization of FE film by producing a ferroelectrically
dead layer. Here, the doping is inhomogeneous among the interface
and the interior of BaTiO3 film. Things will be different for
(NiO$_2$)$^-$/(BaO)$^0$ polarity-discontinuity interfaces in Fig.
4(b), where Fermi energy shifts to the top of valence band of
BaTiO$_3$, and the interfacial BaO layer becomes obviously
metallic, which is induced by the hole doping in the interfacial
BaO layer. If the DOS of the interfacial BaTiO$_3$ layer is
conductive, then we say the interface is metallic, i.e., the
contact between the electrode and BaTiO$_3$ barrier is Ohmic
contact$^{13}$. Similarly, when the DOS of the interfacial
BaTiO$_3$ layer is insulative, then a tunneling barrier is formed
at the interface, which has also been defined in Ref.~13 for
SrRuO$_3$/{\it{n}}-BaTiO$_3$ interface.

With the presence of polarization in Fig.~5 for the superlattice (LaNiO$_3$)$_{10.5}$-(BaTiO$_3$)$_{8.5}$ with the ferroelectric polarization
pointing to the right, an asymmetric distribution of electronic states is found, with the valence band near the left interface upward and the
conduction band near the right interface downward. For the superlattice with (LaO)$^+$/(TiO$_2$)$^0$ interfaces in Fig.~5(a), ferroelectric
polarization controls the accumulation or depletion of electrons near the left and right interfaces, and it is found that the right interface is
ohmic contact, while the left interface turns to Schottky tunnel barrier with the development of ferroelectricity in the BaTiO$_3$ component.
The left and right interfaces are both ohmic contact for PE state as shown in Fig.~4(a). For the case of (NiO$_2$)$^-$/(TiO$_2$)$^0$ interfaces
in Fig.~5(b), the left interface and right interface are always keeping as ohmic contact regardless of the appearance of spontaneous
polarization.

The number of accumulated charges, either the electrons of
occupied Ti $d$ orbital in TiO$_2$ layer or O $p$ orbital in BaO
layer for (LaNiO$_3$)$_{10.5}$-(BaTiO$_3$)$_{8.5}$ superlattice,
is integrated and shown in Fig.~6. For the superlattice with
(LaO)$^+$/(TiO$_2$)$^0$ interfaces, the polar layer of (LaO)$^+$
leads to a four electron-doped BaTiO$_3$ layers at the right
($d_e=4$). For the superlattice with (NiO$_2$)$^-$/(TiO$_2$)$^0$
interface, on the other hand, the holes are restricted at one unit
cell ($d_{hL}=d_{hR}=1$) and the magnitude at both sides is much
larger than the electron doped cases. Obviously, for the strong
itinerant nature of (NiO$_2$)$^-$ layer, the holes are absorbed in
Fermi sea of the LaNiO$_3$ electrodes. These findings are
different from SrTiO$_3$-LaAlO$_3$ heterointerface, where the
holes delocalize into SrTiO$_3$ at p-type interface and electrons
in n-type interface$^{17}$.

In fact, the value $d_e$ and $d_{h}$ are found independent of the
thickness of BaTiO$_3$, i.e., the occurrence of polarization in
BaTiO$_3$ has little influence on the depth of hole-doping. With
the above analysis, the contrast between electron doping and hole
doping is evident in nanoscale ferroelectrics. The electrostatic
potential as demonstrated in Fig. 1(c) is therefore improper for
the LaNiO$_3$-BaTiO$_3$ superlattice with (NiO$_2$)$^-$/(BaO)$^0$
interfaces, where a more symmetric situation is expected. The
polar discontinuity can be neutralized via interface electronic
reconstruction at oxide heterointerface, however, for the
metal-ferroelectric heterostructure only electrons can penetrate
into BaTiO$_3$ component. Such differences also lead to variation
of ferroelectric critical thickness, which is 4 for hole-doped
case and 6 for electron-doped case.\\

In summary, ferroelectricity in LaNiO$_3$-BaTiO$_3$ superlattices with polar discontinuity at the interfaces are investigated from
first-principles. It is found that the polar discontinuity in the (NiO$_2$)$^-$/(TiO$_2$)$^0$ interface is neutralized within the interface, in
contrast to the case of (LaO)$^+$/(TiO$_2$)$^0$ interface, where the screening electron can penetrate into the BaTiO$_3$ layer of four unit
cells. Such differences make the ferroelectricity in the ultrathin films with (NiO$_2$)$^-$/(BaO)$^0$ interface much stronger than that in the
ultrathin films with (LaO)$^+$/(TiO$_2$)$^0$ interfaces, therefore, providing an avenue to realize the interface control of ferroelectricity as
well as functionalities in ferroelectric nanodevices.

\

\textbf{Acknowledgments} Y. Z. Wu thanks Prof.~Valeriy Stepanyuk
for discussions during visiting Max Planck Institute of
Microstructure Physics. S. Ju appreciates Dr. Gaoyang Gou for the
helpful discussion on LaNiO$_3$. This work was supported by the
National Science Foundation of China (Grant Nos.~11274054,
11374220, 11104193, and 11174043), and the 333 project of Jiangsu
Province.

\newpage

\begin{figure}
\vskip -0.9cm
\includegraphics
[width=0.8\textwidth]{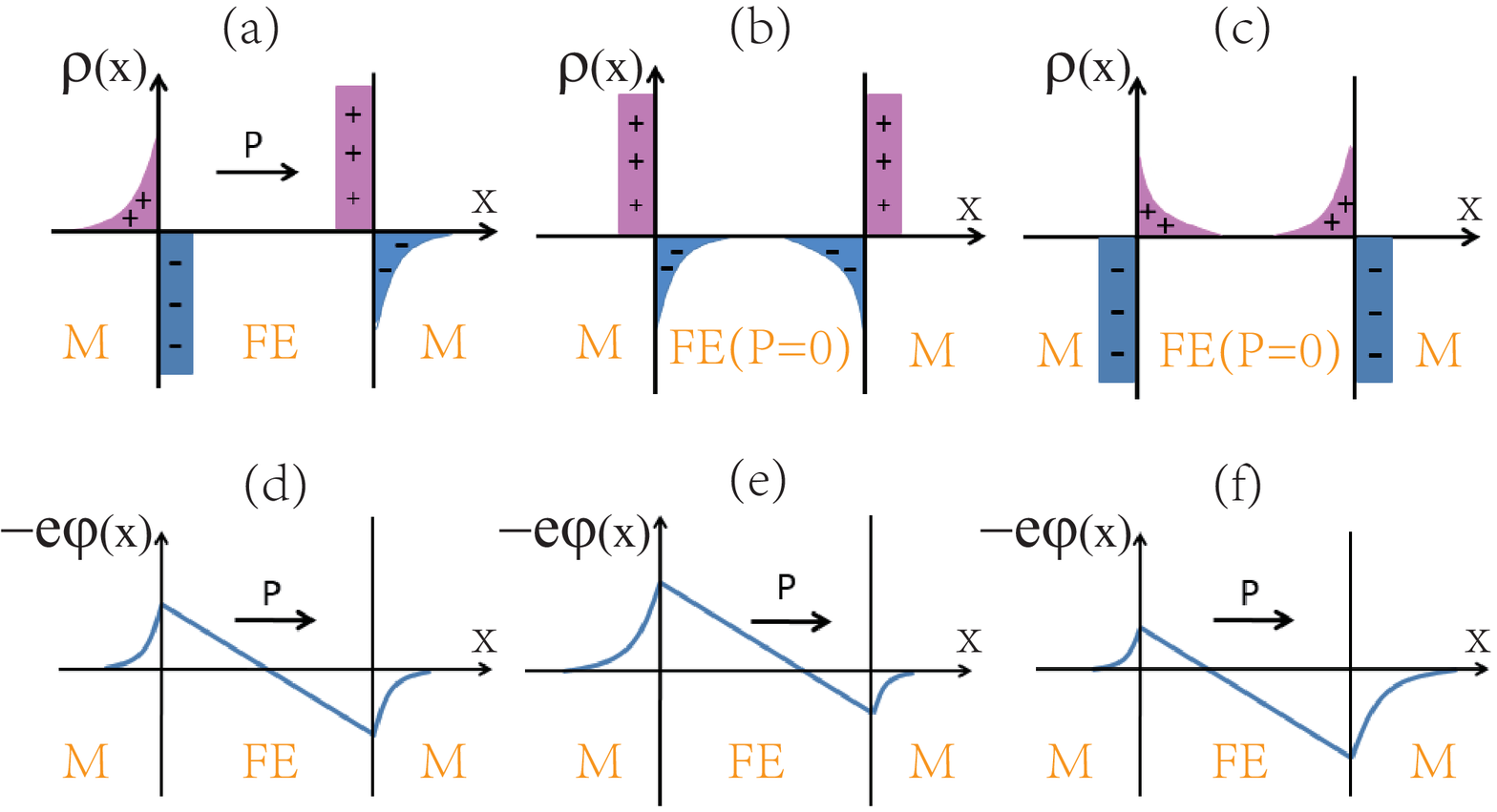} \vskip -0.5cm
\caption{Illustration of charge distribution and potential energy
profile for metal-ferroelectric heterostructures. (a) Electrode of
normal metal. Oxide electrode with polar discontinuity at the
interface with (b) positively charged layer and (c) negatively
charged layer. The potential energy profiles with the inclusion of
spontaneous polarization are shown in (d), (e), and (f),
respectively. In order to show the polar discontinuity and charge
distribution clearly, the barrier polarization are not considered
in (b) and (c), while the spontaneous polarization are included
for the potential energy diagrams in (d), (e), and (f).}
\end{figure}
\

\newpage

\begin{figure}
\vskip -0.9cm
\includegraphics[width=0.8\textwidth]{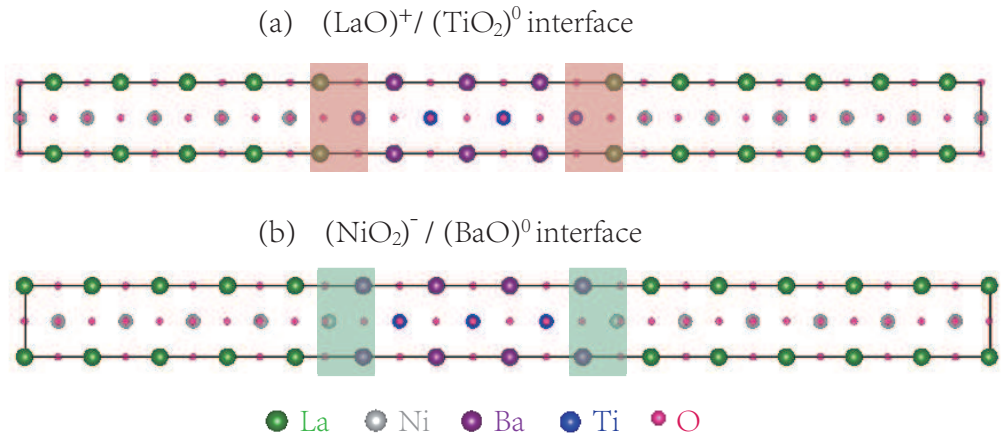}
\vskip -0.5cm \caption{Illustration of LaNiO$_3$-BaTiO$_3$ superlattice structure with (a) (LaO)$^+$/(TiO$_2$)$^0$ interface, and (b)
(NiO$_2$)$^-$/(BaO)$^0$ interface. Here, the thickness of BaTiO$_3$ is 3.5 unit cells.}
\end{figure}
\

\newpage

\begin{figure}
\vskip -0.9cm
\includegraphics[width=0.8\textwidth]{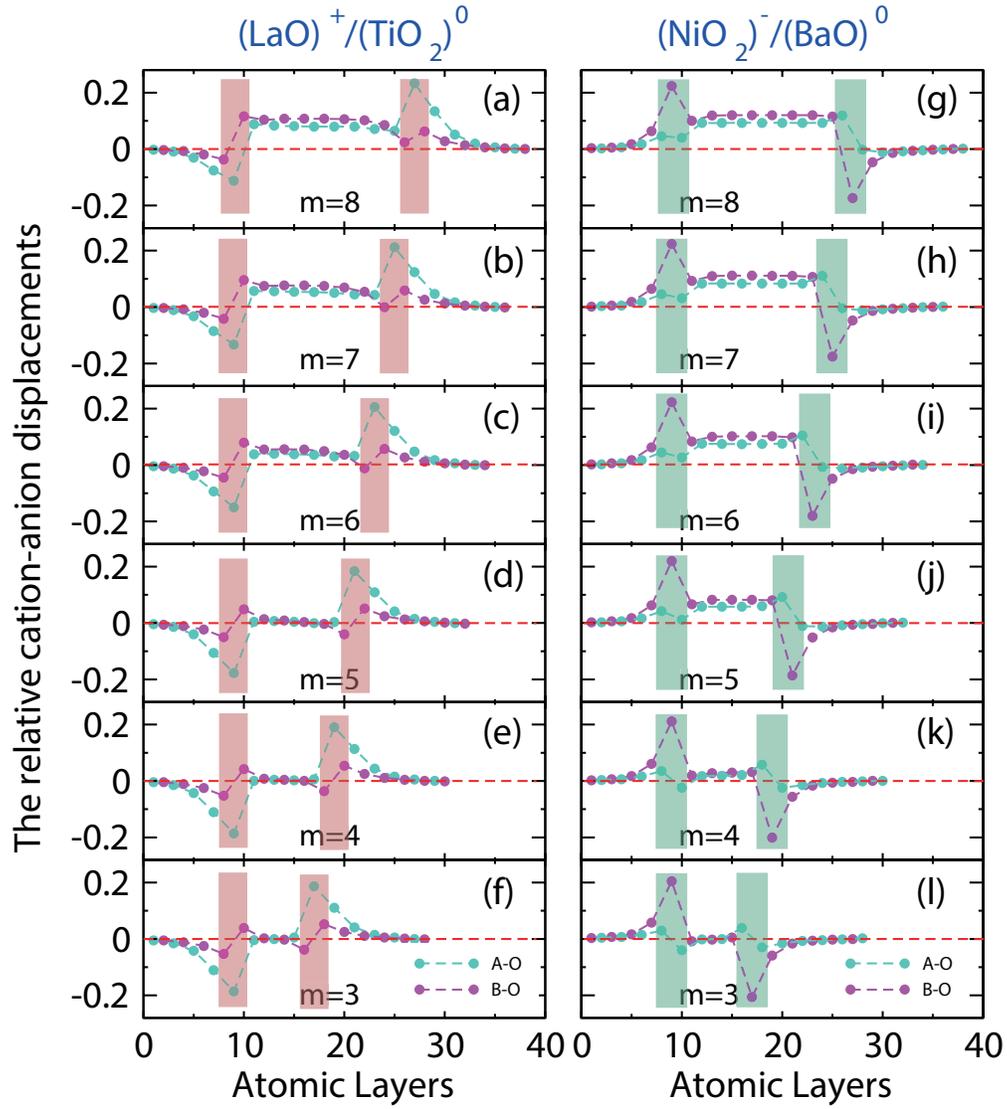}
\vskip -0.5cm \caption{[(a)-(f)] Relative A-O and B-O displacements in the optimized [LaO-(NiO$_2$-LaO)$_{10}$/(TiO$_2$-BaO)$_m$-TiO$_2$]
($m=8, 7, 6, 5, 4, 3$) superlattices. [(g)-(l)] [NiO$_2$-(LaO-NiO$_2$)$_{10}$/(BaO-TiO$_2$)$_m$-BaO] ($m=8, 7, 6, 5, 4, 3$) superlattices.
Shadow regions represent the interfaces.}
\end{figure}
\

\newpage
\begin{figure}
\vskip -0.5cm
\includegraphics[width=0.85\textwidth]{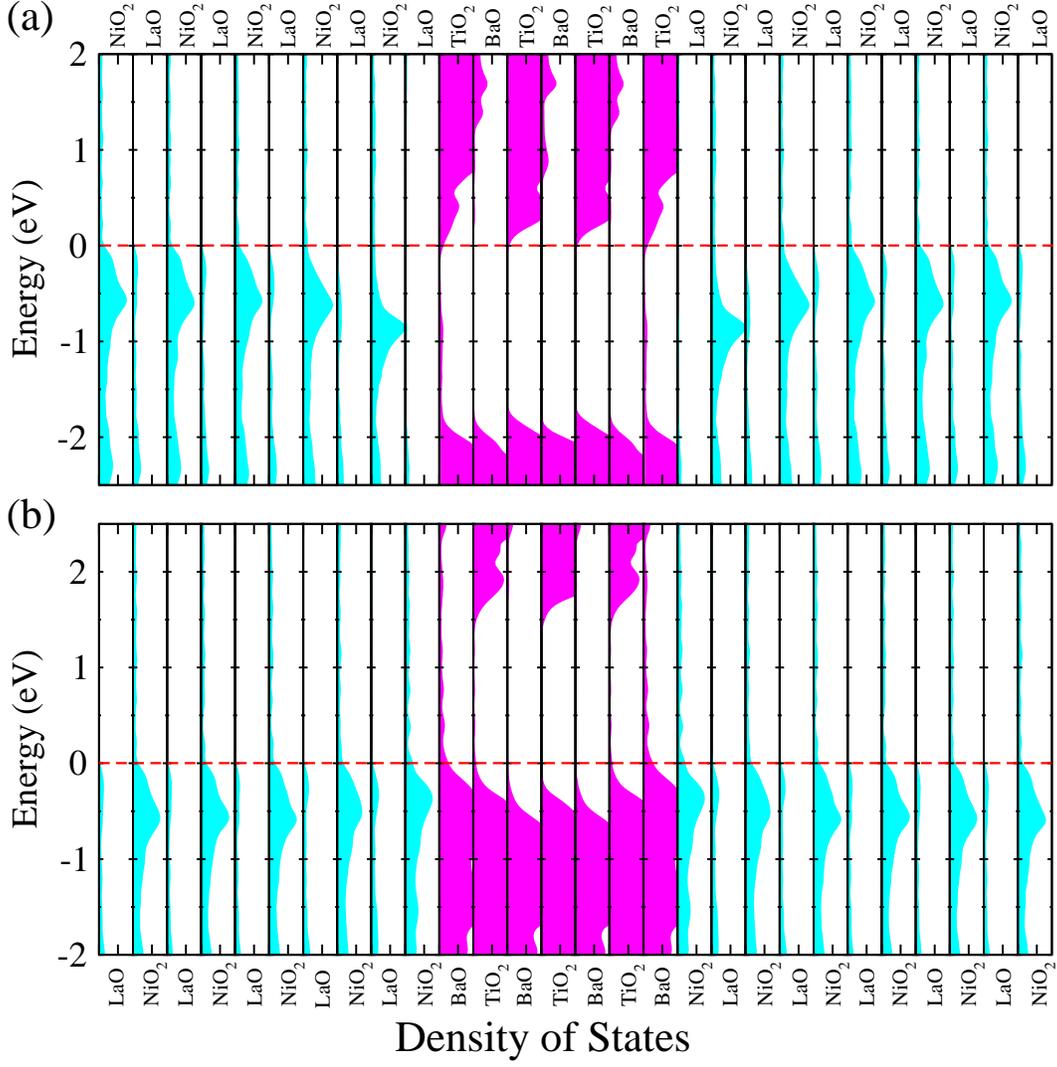}
\vskip 0.5cm
  \caption{Layer-projected density of states in LaNiO$_3$-BaTiO$_3$ superlattices ($m$=3) with (a) (LaO)$^+$/(TiO$_2$)$^0$ interfaces
  and (b) (NiO$_2$)$^-$/(BaO)$^0$ interfaces. The scale of density of states in LaNiO$_3$ is nine times of that in BaTiO$_3$. Here, the component BaTiO$_3$
  stays on paraelectric state.}
\end{figure}
\

\newpage
\begin{figure}
\vskip -0.5cm
\includegraphics[width=0.85\textwidth]{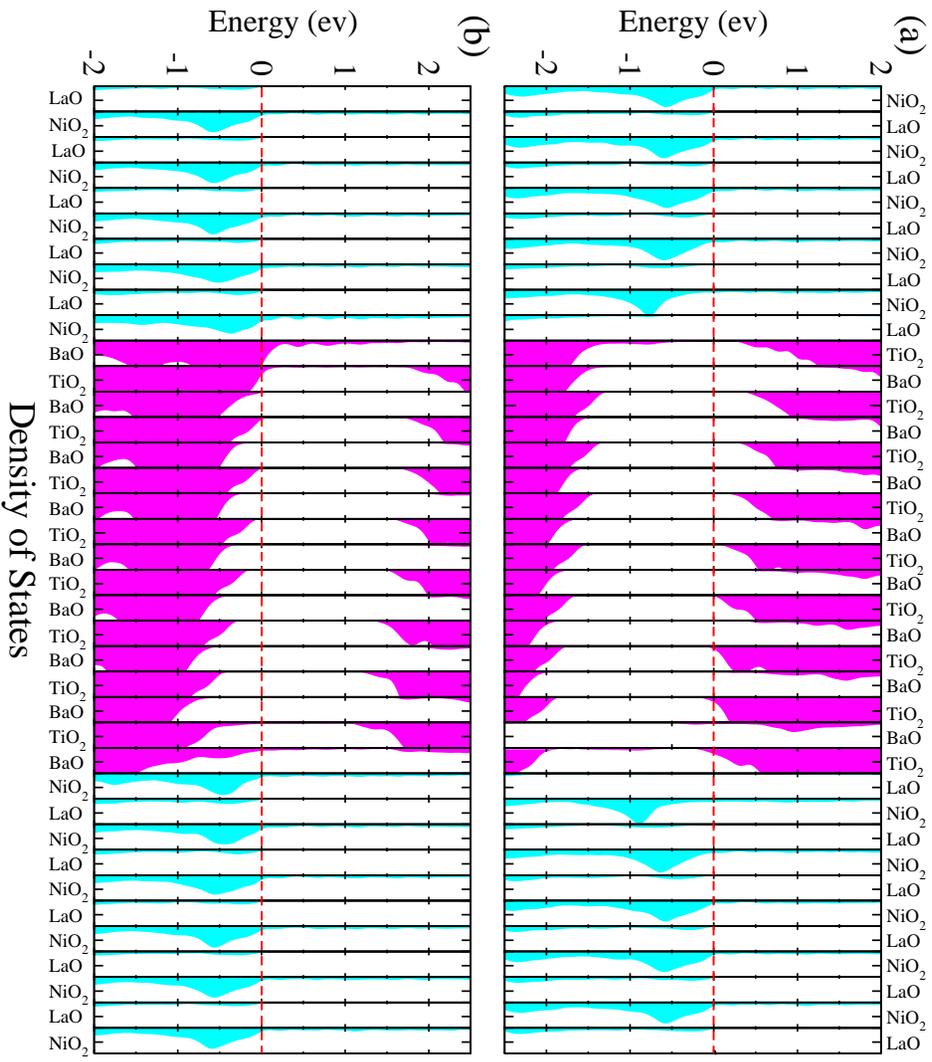}
\vskip 0.5cm
  \caption{Layer-projected density of states in LaNiO$_3$-BaTiO$_3$ superlattices ($m$=8) with (a) (LaO)$^+$/(TiO$_2$)$^0$ interfaces
  and (b) (NiO$_2$)$^-$/(BaO)$^0$ interfaces. The scale of density of states in LaNiO$_3$ is nine times of that in BaTiO$_3$.}
\end{figure}
\

\newpage
\begin{figure}
\vskip -0.9cm
\includegraphics[width=0.66\textwidth]{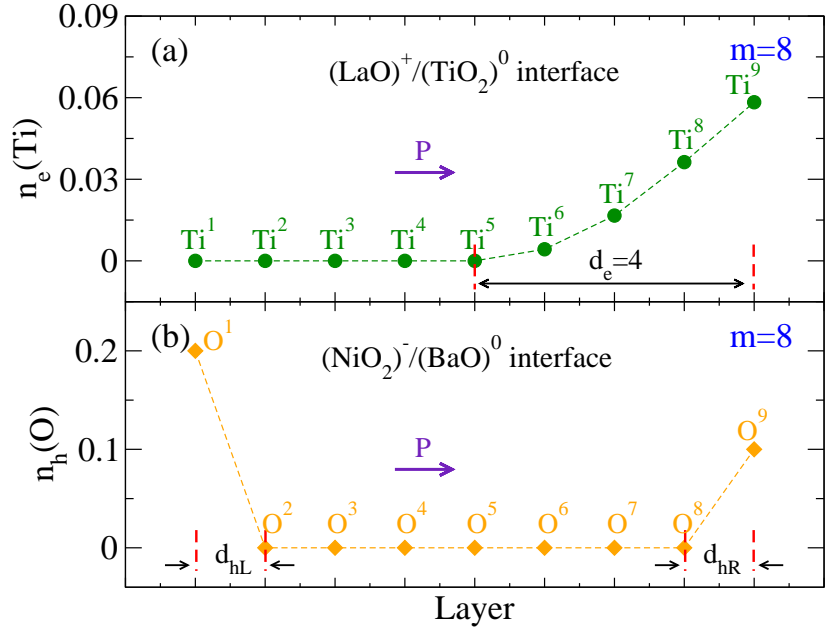}
\vskip -0.5cm \caption{(a) Number of electrons $n_e$ integrated over Ti $d$ orbital in conduction band of each TiO$_2$ monolayer. (b) Number of
holes $n_h$ integrated over O $p$ orbital in the valence band of each BaO monolayer. Here, the thickness of BaTiO$_3$ component is 8.5 unit
cells, and the polarization points to the right.}
\end{figure}
\

\end{document}